# High-speed plasmonic electro-optic beam deflectors

MARTIN THOMASCHEWSKI[1], CHRISTIAN WOLFF, SERGEY I. BOZHEVOLNYI[2]

*Centre for Nano Optics, University of Southern Denmark, Campusvej 55, DK-5230, Odense M, Denmark*
[1]*e-mail:math@mci.sdu.dk,* [2]*e-mail:seib@mci.sdu.dk*

**Abstract:** Highly integrated active nanophotonics addressing both device footprint and operation speed demands is a key enabling technology for the next generation optical networks. Plasmonic systems have proven to be a serious contender to alleviate current performance limitations in electro-optic devices. Here, we demonstrate a plasmonic optical phased array (OPA) consisting of two 10-μm-long plasmonic phase shifters, utilized to control the far-field radiation pattern of two subwavelength-separated emitters for aliasing-free beam steering with an angular range of ±5° and flat frequency response up to 18 GHz (with the potential bandwidth of 1.2 THz). Extreme optical and electrostatic field confinement with great spatial overlap results in high phase modulation efficiency ($V_\pi L = 0.24$ Vcm). The demonstrated approach of using plasmonic lithium niobate technology for optical beam manipulation offers inertia-free, robust, ultra-compact and high-speed beam steering.

## 1. Introduction

Rapid development of large data-handling optical networks necessitates breakthrough concepts and design principles in information and communication technology. Optical modulators, switches and beam-shaping and steering devices are crucial components in modern complex optical control systems for diverse information-processing applications based on optical interconnects. To facilitate their high-speed and low-power operation, it is imperative to increase the efficiency and integration density of dynamic optical components [1]. The ability to dynamically control the phase of photonic channels at GHz frequencies is a crucially important functionality of integrated electro-optic devices such as Mach-Zehnder modulators, directional coupler switches, spatial light modulators or beam deflectors. The most prominent electro-optic material with wide-spread use in the industry is lithium niobate with its birefringence being tunable via the linear electro-optic Pockels effect. This material fulfills eligible material requirements of exhibiting a wide optical transparency (0.35−4.5μm), high electro-optic activity (30 pmV$^{-1}$) which are preserved at harsh environments due to its high Curie temperature (≈ 1200 °C) and excellent chemical and mechanical stability [2]. These properties permit highly reliable and versatile dynamic components. However, current photonic technology utilizes poorly confined metal-diffused optical waveguides, which are limiting the electro-optic interaction resulting in low modulation efficiencies.

Recent progress in high-index contrast lithium niobate waveguides [3–7] resulted in significantly improved device performances regarding modulation efficiency and speed compared to conventional lithium niobate modulators. At the same time, integrated nanophotonics involving plasmonic modes was shown to enable significantly advances in further miniaturization of electro-optic devices [8,9]. Regarding dynamic components, the most efficient modulation performance has been achieved by using plasmonic lithium niobate modulator technology [10]. The high integration capability and the enhanced light-matter interaction of plasmonics has led to a drastically reduced half-wave voltage-length product of 0.21 V·cm, which is the lowest one reported when using the lithium niobate platform. This strong modulation efficiency enables operation on very small device footprints with increased operation bandwidths far beyond 100 GHz. Despite the evidence that the performance of various dynamic components can strongly benefit from plasmonic lithium niobate technology, its implementation remains largely unexplored, lagging behind the recent developments in the photonic lithium niobate technology [3-7].

As an important optical element in photonic network systems, reconfigurable optical beam deflectors allow active control over the shape and direction of radiation. Traditional beam steering approaches such as those employing microelectromechanical system (MEMS) mirrors rely on mechanical movement and have therefore limited scanning speed (≈ few μs), are bulky, consume much energy and are vulnerable in harsh environments [11–14]. Non-inertial beam steering realized by optical phased arrays (OPA) offer a path to reducing the component size and device vulnerability [15–24]. Resembling its radiofrequency counterpart, they consist of an array of coherent electromagnetic emitters. The relative phase between the radiating element is actively tuned, so that the emitted light interferes constructively or destructively, depending on the emission angle and the emitter spacing. A common way to tune the relative phase in integrated OPA systems is by thermo-optic tuning [25–28]. Although the approach of using the strong thermo-optic effects provides a large steering range of more than 30°, devices have remained too large, slow and power-hungry to fulfill the demands of future technology. Independent on the phase-shift mechanism, the majority of present OPAs are based on photonic platforms whose emitters are separated by a distance larger than the wavelength $\lambda_0$ of light. As known from radio waves OPA theory [29], any emitter pitch spacing much greater than half the wavelength in free-space leads to spatial aliasing (presence of grating side lobes) and consequently excessive side lobe radiation that reduces the optical power transmitted in the main steering lobe (See Supplementary Note 3).

Thus, it remains an open research challenge to develop nanophotonic sub-wavelength pitch electro-optic beam deflectors with comprehensive merits such as fast steering speed, wide steering range, high steering resolution, robust design, and compact footprint size.

Here, we describe electrically tunable beam steering, which utilizes a plasmonic platform to enable exceptionally efficient and fast steering in an ultra-compact device footprint. Owing to the Pockels effect in lithium niobate, the phase of highly confined light transmitted through two plasmonic strip waveguides can be continuously tuned by the electrostatic modification of the refractive index in LN. The 10-μm-long waveguides are driving sub-wavelength radiating elements, whose radiation pattern is manipulated by the phase difference between the two channels. The design is robust, and the fabrication - straightforward, with only one lithography step required. This simple and efficient approach combines highly desirable characteristics for various technological beam steering applications, such as Light Detection and Ranging (LiDAR) mapping, free-space communications and spatially resolved optical sensors.

## 2. Results

### 3.1 Design principles and geometry of plasmonic LN beam deflectors

The beam deflector demonstrated in this study is based on two identical subwavelength gold nanowires on bulk lithium niobate (Figure 1). The 400 nm wide plasmonic waveguides support propagation of surface plasmon polariton (SPP) modes which are excited by gold grating couplers with a calculated coupling efficiency of 10 % at the wavelength of $\lambda_0 = 1550$ nm. The grating couplers terminating the plasmonic waveguide serve as emitters of the two-element phased array and are separated by a subwavelength center-to-center distance of 950 nm. The metal nanowires supporting the propagation of plasmonic modes can simultaneously be utilized for introducing an electrostatic field in LN to actively tune the optical signal. By applying a voltage across the wires, the electrostatic field is directed in opposite directions underneath the Au nanowires. Owing to the linear electro-optic effect in LN, the refractive index experiences an opposite modification in response to the electrostatic field. The excellent spatial overlap between optical field and the modification of the refractive index results in a significant phase change with opposite polarities between the optical channels (push-pull phase modulation). This strong electro-optic interaction is characterized by a small half-wave voltage-length product of only 0.24 Vcm for the considered geometry. This quantity can be further reduced by shrinking down the waveguide dimensions, and thereby enhancing the optical confinement and the interaction with the active medium. This comes at the expense of increased propagation losses. In this study, we target a propagation loss of less than 0.4 dB/μm (Supplementary Note 3), so that the total transmission loss of a 10μm-long waveguide does not exceed 4 dB for a waveguide with of 400 nm. A stronger modification of the refractive index for a given bias voltage can also be achieved by reducing the separation between the plasmonic waveguides. However, the operation of the proposed beam deflector requires low crosstalk (−15 dB) between the waveguide channels. Due to a strong optical confinement of the propagating mode in the plasmonic waveguides, a center-to-center separation distance of only 0.9 μm is required to fulfill the condition for low crosstalk (Supplementary Note 1). This separation distance is close to the optimum condition of a half-wavelength spacing between the radiating elements for aliasing-free beam steering [29].

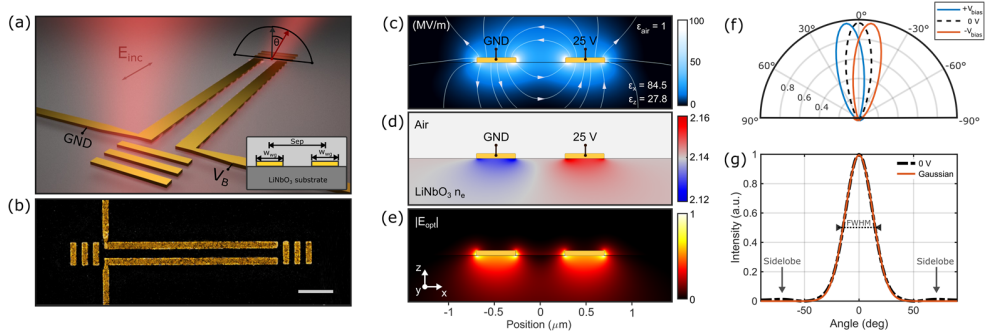

Fig. 1. Integrated electro-optic beam deflector based on lithium niobate (LiNbO$_3$, LN) (a) Conceptual image of beam deflection of the outcoupled optical power by applying a bias voltage across the two nanowire plasmonic waveguides. Cross-section of the two parallel waveguides (Inset). (b) False-colored scanning electron microscope (SEM) image of the investigated plasmonic electro-optic beam deflector. The scale bar represents 2μm. (c) Electrical field distribution and its contours upon applying a voltage of 25 V across the plasmonic nanowires. (d) Electro-optically induced modification of the extraordinary refractive index in lithium niobate. Due to the present electrostatic field, the refractive index is decreased underneath one wire, while it is increased by an equal amount in the opposite wire. (e) The optical field distribution of the supported antisymmetric mode overlaps greatly with the change of the refractive index. (f) Far-field radiation pattern with and without electro-optically induced phase shift of ± 0.4π. (g) Simulated radiation pattern of the unbiased device with a Gaussian fit.

Numerical investigation of the radiation characteristics of the optical field emitted at the waveguide termination were performed by using the near-to-far-field transformation [30]. The near-field distribution and phase of the individual channels were calculated from the plasmonic mode profiles and the electro-optic interaction along the phase shifter (Supplementary Note 2). The waveguide termination results in scattering of the phase-detuned near-fields out of the surface plane, thereby forming a specific far-field radiation pattern through interference. The shape and direction of the radiated optical beam is mainly defined by the waveguides cross-section geometry and the phase difference between the emitters. The steering angle θ is equal to the angle of the emitted phase front. Therefore, reducing the separation distance steers the optical beam to larger angles for the same near-field phase difference. As stated above, a smaller separation distance also implies stronger electro-optic interaction, which additionally contributes to increased steering angles. Due to the subwavelength separation distance between the scattering centers in our OPA, aliasing is significantly suppressed with an estimated sidelobe level of only -17dB for the considered geometry (Figure 1 g). Therefore, a large part of the energy in the main lobe can be efficiently deflected. For the unbiased device the emitters are in phase resulting in an undeflected main lobe directed towards the surface normal. Depending on the polarity of the introduced relative phase difference, the beam experiences either a positive or negative azimuth deflection.

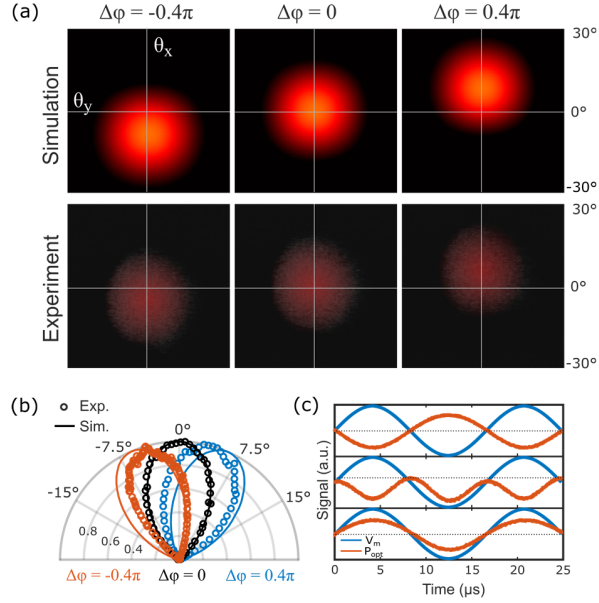

Fig. 2. (a) Simulated and experimentally obtained Fourier images of the beam steering showing the one-dimensional angular deflection of the beam from the surface normal shown in the origin. (b) Integrated intensities along the of the radiation profiles along the horizontal axis reveal the far-field radiation pattern Intensity modulation detected with an angular filter in the Fourier plane positioned at θ < 0 ° (top), θ = 0 ° (middle) and θ > 0 ° (bottom).

### 3.2 Demonstration and characterization of dynamic plasmonic beam steering

With the above theoretical framework, we now move on to the experimental realization of the plasmonic electro-optic beam steering. The proposed device is fabricated by standard EBL on bulk LN substrate followed by gold deposition and lift-off (See methods for details). A focused linearly polarized laser beam with linear polarization is positioned onto the coupler section of the device at normal incidence. Surface plasmon polaritons are launched into the 10-μm-long phase shifters. To characterize the actively-tunable radiation pattern, an infrared CCD camera is used to display the far-field angular distribution of the scattered optical signal at the output grating. At zero bias, the radiation lobe is directed to the normal direction due to in-phase radiation of the emitters. When a bias signal is applied, beam steering occurs. The polarity of the applied voltage, i.e. the sign of the introduced phase shift, determines hereby the direction of steering. Our NFFT simulations reproduce the experimentally obtained Fourier image with regard to the observed steering angle and the beam divergence (Figure 2). By fitting the maxima and width of a Gaussian function to the far-field scattering images, the steering angles and beam divergence can be determined for each applied voltage. The largest drive voltage without permanent device breakdown translates to an electro-optically introduced phase shift of ±0.4π. Therefore, the corresponding steering range is limited to $\theta = \pm 5°$ with a tuning rate of $\Delta\theta/\Delta V = 87$ μrad/V. A relatively large FWHM divergence of 25° is experimentally reported, which is naturally limited by the small size of the nanophotonic two-element phased array. Smaller divergence can be achieved by expanding the emitter size and by increasing the number of emitters, which hinders efficient, compact and fast operation. Dynamic beam steering induced by alternating voltage is captured by placing a high-speed photodiode in combination with an angular circular filter ($\Delta\theta_x \ll \sigma$) in the Fourier plane. As the filter is placed off-center, either collecting only the angular components $\theta > 0°$ or $\theta < 0°$, the detected power signal alternates in-phase or 180° out-of-phase with the driving signal, respectively. When the filter is centered around $\theta = 0°$, the modulated signal follows the second harmonic of the driving signal with the maximum output signal at zero volts, as is to be expected due to symmetry. The experimentally observed voltage-dependent beam deflection agrees well with the theoretical analysis (Figure 3a). Though the deflection angles a slightly smaller (≈ 10 %) than predicted, continuous and quasi-linear steering between −5° and +5° is observed.

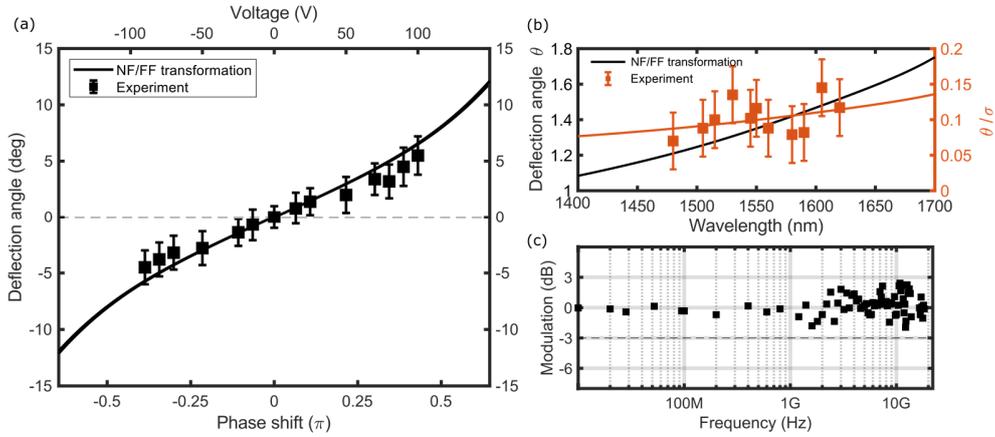

Fig. 3. (a) Beam deflector as a function of the electro-optically induced phase shift between the two radiating emitters. (b) Wavelength-dependent emission characteristics regarding the deflection angle θ and the deflection-divergence ration θ/ρ. (c) Measured response of electro-optic steering as a function of the applied RF signal frequency.

Due to the large transparency of lithium niobate and non-resonant electro-optic phase modulation, spectrally broadband operation of our beam steering device is expected. The dispersion of the modulation efficiency of the plasmonic phase shifters and the wavelength-dependent emission pattern result in distortion of the deflection angle and the beam divergence (Figure 3b). As the deflection angle increases with larger wavelengths, the beam divergence becomes broader. This trade-off is represented by the deflection-divergence ration θ/σ which is observed to be nearly constant over the investigated wavelength range.

One of the most important aspects for current optical phased array technology is its scalability and speed. Due to a record-small device length of only 10 μm, our device exhibits an electrical capacitance of only $C = 2.4$ fF. In combination with the extremely short response times of the utilized Pockels effect, ultra-fast steering operation is expected. We characterized the frequency response of optical beam steering in our device and observed a flat response from 10 Mhz until 18 GHz. The maximum measurable steering speed is only limited by our electrical test equipment. The calculated modulation cutoff frequency is 1.2 THz at $R = 50\ \Omega$ resistive load ($f_c = 1/[2\pi RC]$), therefore providing a possible route for terahertz applications in high-speed communication and sensors.

## 3. Discussion

We report ultra-fast (with 18 GHz demonstrated and 1.2 THz expected bandwidths) optical beam steering across a 10° field of view with a 10-μm long two-element plasmonic phased array. The integrated plasmonic phase shifter exhibits a half-wave voltage length product of only 0.24 V·cm, to date the highest efficiency for electro-optic beam steering via the Pockels effect. Owing to the subwavelength element pitch between the emitters of the OPA, aliasing-free steering with > 15 dB sidelobe suppression is achieved. Numerical simulations support the experimental findings of this work and provide insights into the geometrical dependence on the device performance regarding steering range and divergence. The demonstrated lithium niobate plasmonic approach for optical beam steering could serve as the base for more robust, compact and faster OPA designs utilized in a wide range of applications such as free-space optical communication, laser-radar sensing systems and LiDAR systems.

## 4. Methods

*4.1 Demonstration and characterization of dynamic plasmonic beam steering*

Numerical simulations of the electro-optic response of the phase shifter were carried out using a commercial finite element method solver (Comsol Multiphysics 5.2). For the electrostatic simulations, the waveguide cross-section is modeled with the relative permittivity $\varepsilon_{\text{air}} = 1$ of air and the strain-free static relative permittivity tensor of lithium niobate ($\varepsilon_{xx} = \varepsilon_{yy} = 27.8$, $\varepsilon_{zz} = 84.5$) taken from Jazbinsek et al. [31]. The boundaries of the two gold electrodes are set to ground and $V_{\text{bias}}$, respectively, to calculate the electric field distribution E(x,z). The spatially-dependent modification of the refractive index in LN is then calculated by using the Pockels coefficients from Jazbinsek et al. [31] (considering the largest diagonal terms, i.e., $\Delta n_{ii} = -0.5 r_{iiz} n_{ii}^3 E_z$, with $r_{xxz} = r_{yyz} = 10.12$ pm/V and $r_{zzz} = 31.45$ pm/V). The analysis of the plasmonic modes is conducted by considering the electro-optically modified distribution of the refractive index of LN, while the unmodified refractive indices from Au and LN are taken from Johnson and Christy [32] and Zelmon et al. [33], ($n_{xx} = n_{yy} = n_{\text{o}} = 2.211$, $n_{zz} = n_{\text{e}} = 2.138$ at $\lambda_0 = 1550$ nm). We apply scattering-boundary conditions in combination with a perfectly matched layer to calculate the phase mismatch at the output port of each individual waveguide as a function of the applied voltage. The total far-field emission profile is calculated from the lateral near-field mode distribution (Supplementary Note 2) in combination with the near-to-far field transformation (NFFT) approach. A Gaussian distribution $f_G = A_1 \exp(-(\theta_x - \theta)^2/2\sigma^2)$ is fitted to the main lobe of the far-field radiation pattern to determine the deflection angle and the divergence.

*4.2 Sample fabrication*

The plasmonic beam steering elements are fabricated on commercially available z-cut lithium niobate substrates by electron beam lithography (using a scanning electron microscope JEOL JSM-6490LV with an acceleration voltage of 30 keV) in spin-coated 200-nm-thick PMMA positive resist and 20-nm-thick Al layer, which serves as a metallic charge dissipation layer during the writing (electron doses varying between 200 and 250 μCcm$^2$). After resist development, the devices are formed by depositing a 4 nm titanium adhesion layer and a 50 nm gold layer by thermal evaporation and sub-sequent 12h lift-off in acetone. To reduce electron beam writing time, RF feed lines are patterned beforehand on the LN chip by shadow masks metal deposition (5 nm Ti / 100 nm Au).

**Funding.** The authors acknowledge funding support from the VKR Foundation (Award in Technical and Natural Sciences 2019) and the VILLUM FONDEN (Grant 16498). C. W. acknowledges funding from a MULTIPLY fellowship under the Marie Skłodowska-Curie COFUND Action (grant agreement No. 713694)

**Data availability.** Data underlying the results presented in this paper are not publicly available at this time but may be obtained from the authors upon reasonable request.

**Supplemental document.** See Supplement 1 for supporting content.